\documentstyle[floats,aps]{revtex}
\begin{document}
\input epsf

\font\tenmib=cmmib10
\font\eightmib=cmmib10 scaled 800
\font\sixmib=cmmib10 scaled 667
\newfam\mibfam
\textfont\mibfam=\tenmib
\scriptfont\mibfam=\eightmib

\draft

\twocolumn[\hsize\textwidth\columnwidth\hsize\csname  
@twocolumnfalse\endcsname
\title{Phase separation in double exchange systems}

\author{Daniel P. Arovas$^1$, Guillermo
G\'omez-Santos$^2$ and Francisco Guinea$^{3}$}
\address{$^1$Department of Physics, University of California at San Diego,
 La Jolla CA 92093\\
$^2$ Departamento de F{\'\i}sica de la Materia Condensada
and Instituto Nicol\'as Cabrera, 
Universidad Aut\'onoma de Madrid. Cantoblanco. E-28049 Madrid , Spain\\
$^3$Instituto de Ciencia de Materiales, CSIC,
Cantoblanco, 28049 Madrid, Spain}

\date{May 29, 1998}

\maketitle

\begin{abstract}
Ferromagnetic systems described by the double exchange model
are investigated.  At temperatures close to the Curie temperature,
and for a wide range of doping levels, the system is unstable
toward phase separation.  The chemical potential decreases
upon increasing doping, due to the significant dependence of
the bandwidth on the number of carriers.  The reduction of the
electronic bandwidth by spin disorder leads to an enormously
enhanced thermopower which peaks near $T_{\rm c}$, with a sign
opposite that predicted by a rigid band model.
\end{abstract}

\vskip2pc]

\narrowtext

Doped manganese oxides exhibit many unusual features, most notably
the eponymous phenomenon of colossal magnetoresistance (CMR)
\cite{wk55,g63,cvm97}, and a phase transition from a high temperature
paramagnetic insulator to a low temperature ferromagnetic metal.
The Mn $3d$ states split into a lower  $t_{2g}$\ triplet, forming an
$S= \frac{3}{2}$ core spin, and an upper  $e_g$\ doublet, the conduction band.
This physics is described by the double exchange model \cite{z51}:
the conduction electrons which hop throughout the lattice are
ferromagnetically coupled to the local core spins because of Hund's
rules.  Typical values for the  $e_g$\ hybridization $t$ and intra-atomic
exchange $J_ {\rm H}$ are $t\sim 0.1\,$eV and $J_ {\rm H}\sim 1-3\,$eV.
Additionally, one may choose to incorporate other terms, most notably
Heisenberg couplings between neighboring core spins (due to superexchange)
and Jahn-Teller (JT) phonons which break the degeneracy of the $e_g$\ level.

The simplest model, however, is one in which the degeneracy of
the $e_g$\ orbital is simply ignored
(possibly due to a cooperative JT
distortion), and $J_ {\rm H}$ is set to infinity.  In this limit, the spin of
each  $e_g$\ electron must agree with that of its $t_{2g}$\ core, and
we may write $e ^\dagger_{i\sigma}=
c ^\dagger_i\,z ^{\vphantom{\dagger}}_{i\sigma}$, where
$z ^{\vphantom{\dagger}}_{i\sigma}$ is the spinor
describing the orientation of the core at site $i$:
$z ^{\vphantom{\dagger}}_{i \uparrow}=\cos( \frac{1}{2}\theta_i)$,
$z ^{\vphantom{\dagger}}_{i \downarrow}=
\sin( \frac{1}{2}\theta_i)\,\exp(-i\phi_i)$.
The Hamiltonian is then
\begin{equation}
{\cal H} = -t\sum_{\langle ij\rangle,\sigma}
\left[ z ^{\vphantom{\dagger}}_{i\sigma}
{\bar z} ^{\vphantom{\dagger}}_{j\sigma}\,
c ^\dagger_i c ^{\vphantom{\dagger}}_j + {\rm H.c.}\right]\ .
\label{hamil}
\end{equation}
The hopping is maximized when neighboring core spins are parallel.
The ferromagnetic interaction induced in this way describes qualitatively
the physics of the manganites.  The Curie temperature increases
with the number of carriers, and, in the paramagnetic phase,
the tendency towards localization is enhanced.
On the other hand, it has been extensively argued that 
the double exchange model does not suffice to describe
the insulating behavior at high temperatures \cite{mls95}. 
A disordered distribution of spin gives rise to off-diagonal disorder,
which is in a sense weaker than the more conventional diagonal disorder,
and most of the states in the band are delocalized\cite{rzb96}
(see, however, the comments of \cite{v96}).  The  $e_g$\ bandwidth remains
finite and is unremarkable as one passes through the Curie temperature
$T_ {\rm c}$ \cite{cb98}.  Scattering by magnetic
fluctuations near $T_ {\rm c}$
also is insufficient to explain the insulating regime\cite{mls95}.
A variety of phenomena, including variable range hopping
\cite{cvr95}, formation of spin polarons \cite{ti97}, and the dynamic
Jahn-Teller effect \cite{msm96} have been invoked in an attempt to
understand the insulating paramagnetic phase.

Here we will show that the double exchange model
(\ref{hamil}) is unstable toward charge segregation at low to moderate
doping and temperatures near $T_ {\rm c}$.  This phenomenon arises due to
the coupling between magnetic fluctuations and the electronic
chemical potential.  Exchange interactions between core spins tend
to suppress this effect, but do not completely eliminate it.
The Coulomb interaction will suppress complete phase separation of
charge carriers, resulting in microscopic domains of different charge
density.  This inhomogeneity in turn enhances localization of electrons.
A small ferromagnetic region of high density, surrounded by a low density
paramagnetic background, is indistinguishable from the spin
polaron picture discussed in the literature.

To see why phase separation should occur in (\ref{hamil}), let us
reexamine de Gennes' mean field approach \cite{g60}.  The quantity
$ {\overline W}\equiv\sqrt{\langle |z ^{\vphantom{\dagger}}_{i\sigma}
{\bar z} ^{\vphantom{\dagger}}_{j\sigma}|^2\rangle}
=\langle\cos^2( \frac{1}{2} \vartheta_{ij})\rangle^{1/2}$
is a ``spin reduction factor'' which
multiplies the fermion hopping $t$, compressing the bare ($ {\overline W}=1$)
dispersion $E$ to $ {\overline W} E$.  At zero
temperature, states with $ E <
E ^{\vphantom{\dagger}}_{\scriptscriptstyle {\rm F}}$
are filled, and $\mu= {\overline W}
E ^{\vphantom{\dagger}}_{\scriptscriptstyle {\rm F}}$.
The density of states $D( E)$ and doping
$x\equiv \frac{1}{2}+\delta$ determine $\mu$ via
$\delta=\int_0^{E^{\vphantom{\dagger}}_{\scriptscriptstyle{\rm F}}}
\! d E\, D( E)$.  Thus,
\begin{equation}
{ \partial\mu\over\partial\delta}={{\overline W}\over D(\mu/ {\overline W})}
+{\mu\,\over {\overline W}}\,{ {\partial} {\overline W}\over
{\partial}\delta}\ .
\label{dmdd}
\end{equation}
When $ {\partial}\mu/ {\partial}\delta<0$, the system can lower
its energy through
phase separation.  Now at zero temperature we should expect perfect
order in the spins, hence $ {\overline W}=1$ and the compressibility
is positive.  At finite temperature, however, we expect
$\mu\, {\partial} {\overline W}/ {\partial}\delta<0$
since increased carrier density $|\delta|$ enhances the fermion kinetic
energy, which serves as an exchange coupling for the spins (this is true
whether the carriers are electrons or holes).  If we assume the
fermions to be degenerate, the condition for negative compressibility
is $\mu D(\mu/ {\overline W})\, {\partial}{\overline W}/
{\partial}\delta<-{\overline W}^2$.  To satisfy this relation
requires both finite $T$ and low to intermediate $|\delta|$
(note $\mu=0$ for $\delta=0$, assuming a symmetric band).
As we shall see, this behavior may persist above
the Curie temperature as well.  At sufficiently large $T$, however,
the fermions are nondegenerate and the chemical potential dominates
the bandwidth, giving $\delta= \frac{1}{2}\tanh(\mu/ 2
k_{ \scriptscriptstyle{\rm B}}T)$ and $\partial\mu/\partial\delta =
4 k_{ \scriptscriptstyle {\rm B}}T/(1-4\delta^2)>0$.

This picture also leads to an anomalous thermopower \cite{am76}.
At low temperatures the variation of the chemical potential with $T$
arises from two sources: the Fermi distribution ({\it e.g.\/} the Sommerfeld
expansion) and the band narrowing ${\overline W}$.  Thus one obtains
\begin{equation}
{\partial\mu\over\partial T}=\mu_\circ\,{\partial{\overline W}\over\partial T}
-{\pi^2\over 3}\, {D'(\mu_\circ)\over D(\mu_\circ)}\,
k_{\scriptscriptstyle\rm B}^2 T +\ldots
\end{equation}
For models where $D'(E)/E >0$, such as the elliptical density of states,
the first term, arising from the band narrowing, opposes the second.
As we shall see for all but the lowest and highest temperatures it is
the first term which dominates, and the critical behavior of ${\overline W}$
in the vicinity of $T_{\rm c}$ leads to a peak in the thermopower which
is several orders of magnitude larger than that usually encountered in
metals.

Adding purely ferromagnetic superexchange between the core spins
reduces the dependence of $W$ on $\delta$ and thereby opposes phase
separation.  In  La$_{1-x}$A$_x$MnO$_3$,
however, the interplane superexchange is
antiferromagnetic for small $x$, and its competition with double exchange
leads to a canted structure \cite{g60}.  The bandwidth then depends on
the canting angle and increases with increasing number of carriers.
This property leads to phase separation even at $T=0$ \cite{n96,dym97,ag97}.

There is a subtlety in the mean field theory surrounding which
density of states $D(E)$ one should use.  Integrating out
the fermions from (\ref{hamil}) gives a free energy
\begin{equation}
F=\sum_{p=1}^\infty {(-1)^p\over p!}\,I^{(p)}(-\mu)
\mathop{\rm Tr}_{\{ {\hat {\fam\mibfam\tenmib\Omega}}_i\}}\! H^p
-TS[\{ {\hat {\fam\mibfam\tenmib\Omega}}_i\}]
\label{free}
\end{equation}
where $I(\varepsilon)=- k_{ \scriptscriptstyle {\rm B}}T
\ln(1+e^{-\varepsilon/ k_{ \scriptscriptstyle {\rm B}}T})$ and $H_{ij}=
z ^{\vphantom{\dagger}}_{i\sigma} {\bar z} ^{\vphantom{\dagger}}_{j\sigma}
\,T_{ij}$ with $T_{ij}=t$ if $i$ and $j$
are nearest neighbors and $0$ otherwise.  The entropy
$S[\{{\hat{\fam\mibfam\tenmib\Omega}}_i\}]=- k_{ \scriptscriptstyle {\rm B}}
\mathop{\rm Tr} \varrho\,\ln \varrho$ and the trace in (\ref{free})
are computed from a trial density matrix
$ \varrho$ for the spins.  Assuming an uncorrelated
$ \varrho=\prod_i P({\hat{\fam\mibfam\tenmib\Omega}}_i)$ with
$P({\hat{\fam\mibfam\tenmib\Omega}})\propto \exp(Q\Omega_z)$,
we have that $\langle \exp(i\phi_i)\rangle=0$, hence in the locator
expansion of $\mathop{\rm Tr}_{\{ {\hat {\fam\mibfam\eightmib\Omega}}_i\}}H^p$,
many of the $2^p$ terms associated with a given length $p$ path average
to zero.  These cancellations are avoided for paths which retrace themselves,
for which each $z ^{\vphantom{\dagger}}_{i\sigma}
{\bar z} ^{\vphantom{\dagger}}_{j\sigma}$ term has a
$ {\bar z} ^{\vphantom{\dagger}}_{i\sigma}z ^{\vphantom{\dagger}}_{j\sigma}$
mate.  Thus, in the vicinity of $T_{\rm c}$ where magnetic
fluctuations are significant, it is better to
use the retraced path approximation \cite{br70},
\begin{equation}
D( \gamma)={2\over\pi}{ \sqrt{1- \gamma^2}\over
{{z\over z-1}-{4\over z} \gamma^2} }\,
\end{equation}
where $z$ is the lattice coordination number, in one's mean field
calculations.  $D( \gamma)$, given above 
with $\gamma=-E/B$ in units of the bare
half-bandwidth $B=2(z-1)^{1/2}t$, interpolates between the exact
one-dimensional tight binding density of states at $z=2$ to a semi-elliptic
form at $z=\infty$.  The hopping is then modulated by $ {\overline W}$, the
root-mean-square average of
$z ^{\vphantom{\dagger}}_{i\sigma} {\bar z} ^{\vphantom{\dagger}}_{j\sigma}$.

The mean field equations of de Gennes' model are
\begin{eqnarray}
\delta&=&{1\over 2}\int_{-1}^1\!\!\!\! d \gamma\, D( \gamma)\,\tanh
\left({\alpha +{\overline W}  \gamma\over 2\Theta}\right)\label{dg1}\\
2\Theta\,{ {\overline W}Q\over M}&=&
{1\over 2}\int_{-1}^1\!\!\!\! d \gamma\, \gamma\, D( \gamma)\,
\tanh\left({\alpha +{\overline W}\gamma \over 2\Theta}\right)
\label{dg2}
\end{eqnarray}
where $M= \,{\rm ctnh\,}(Q)-Q^{-1}$ is the magnetization,
and ${\overline W}=\sqrt{ \frac{1}{2}(1+M^2)}$.
The temperature and chemical potential are scaled
by $\alpha\equiv\mu/B$ and $\Theta\equiv k_{ \scriptscriptstyle {\rm B}}T/B$.

\begin{figure} [!t]
\centering
\leavevmode
\epsfxsize=8cm
\epsfysize=8cm
\epsfbox[18 144 592 718] {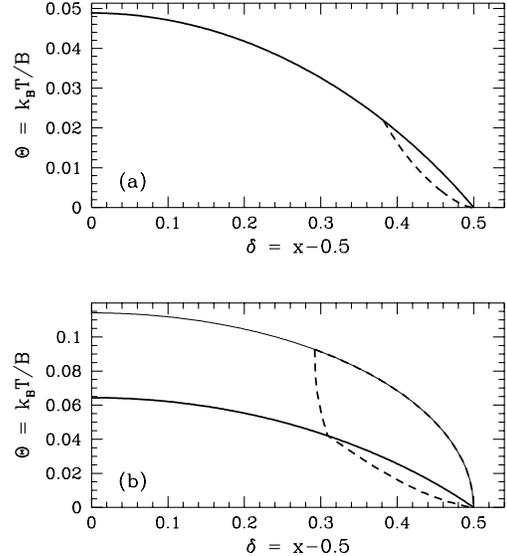}
\caption[]
{\label{fig1}
Phase diagrams (a) for the mean field theory of eqns. (\ref{dg1},\ref{dg2}),
and (b) for that of eqns. (\ref{hmf},\ref{sbmf}), both computed using an
elliptic density of states.  The thick solid line is $\Theta_{\rm c}(\delta)$.
The thick dashed line is $\Theta_{\rm s}(\delta)$, the boundary of the
region of phase separation.  In (b), the thin solid line denotes
$\Theta_*(\delta)$, the temperature where the bandwidth collapses.}
\end{figure}

The mean field solution exhibits a Curie temperature
$\Theta_ {\rm c} (\delta)$.
With $t\approx 100\,$meV \cite{ps96}, the cubic lattice model
has $B=2\sqrt{5}\,t=5200\,$K.  We find a maximum $\Theta_{\rm c}=0.0545$
at $\delta=0$, corresponding to a Curie temperature of $T_{\rm c}=280\,$K,
very much consistent with experimental values.
$M$ vanishes at $\Theta_ {\rm c}$, and for $\Theta\ge\Theta_ {\rm c}$ the
bandwidth is finite, reduced by the spin factor
$ {\overline W}=\frac{1}{\sqrt{2}}$.
The compressibility $\kappa^{-1}\equiv{\partial}\alpha/{\partial}\delta$
and thermopower ${\cal S}= {\partial}\alpha/ {\partial}\Theta$
are discontinuous at the transition.  ${\cal S}$ peaks near $T_{\rm c}$
with an anomalously large value; we shall return to this point below.
For the $z=\infty$ case, we find a crescent region of phase separation
$\Theta\in[\Theta_ {\rm s} (\delta),\Theta_ {\rm c} (\delta)]$ extending from
$\delta=0.382$ to $\delta= \frac{1}{2}$, in qualitative agreement with our
earlier discussion.  In Fig. \ref{fig1}(a) we plot the mean field phase
diagram for the limiting case $z=\infty$.  (As one approaches the
one-dimensional limit $z=2$, the region of phase separation is enlarged.)
For $z>2$ we find $\Theta_ {\rm c} \propto {\overline x}$ and
$\Theta_ {\rm s}\propto {\overline x}^{5/3}$ as
${\overline x}= \frac{1}{2}-\delta \to 0$.
(The model is of course invariant under $x\leftrightarrow
{\overline x}$.)

An unphysical aspect of the preceding mean field analysis is that
${\overline W}$, which should be determined by local spin correlations, is
tied to the magnetization $M$.  Initially ${\overline W}$ decreases linearly
with $\Theta$, leading to artifactual behavior in the low temperature
magnetization and thermopower.  The root of the problem is the classical
treatment of the core spins.  At temperatures $\Theta < {\cal O}(1/S)$,
one must treat the spins quantum mechanically.  We should then expect,
from spin wave theory, that ${\overline W}=1-{\cal O}(\Theta^{5/2})$.

A somewhat more sophisticated mean field theory can be implemented using the
Schwinger boson \cite{aa88} method.  A nondynamical field $\lambda_i$
enforces the constraint ${\bar z}^{\vphantom{\dagger}}_{i\sigma}
z^{\vphantom{\dagger}}_{i\sigma}~=~1$ at every site.  The core spins are
quantized according to $[z ^{\vphantom{\dagger}}_{i\sigma},
{\bar z} ^{\vphantom{\dagger}}_{j\sigma'}]=
\delta_{ij}\delta_{\sigma\sigma'}/2S$.
The mean field Hamiltonian is obtained through a Hartree decoupling of
the bosonic $z ^{\vphantom{\dagger}}_{i\sigma}
{\bar z} ^{\vphantom{\dagger}}_{j\sigma}$ and
fermionic $c ^\dagger_i c ^{\vphantom{\dagger}}_j$
hopping terms:
\begin{eqnarray}
{\cal H}_{\scriptscriptstyle{\rm MF}}&=&N(ztWK - \lambda) -\mu\sum_i
c ^\dagger_i c ^{\vphantom{\dagger}}_i
+\lambda\sum_{i,\sigma}  {\bar z} ^{\vphantom{\dagger}}_{i\sigma}
z ^{\vphantom{\dagger}}_{i\sigma}
\label{hmf}\\
&&-tW\sum_ {\langle ij\rangle} (c ^\dagger_i c ^{\vphantom{\dagger}}_j
+ c ^\dagger_j c ^{\vphantom{\dagger}}_i) -tK
\sum_{ {\langle ij\rangle},\sigma}
({\bar z} ^{\vphantom{\dagger}}_{i\sigma} z ^{\vphantom{\dagger}}_{j\sigma}
+ {\bar z} ^{\vphantom{\dagger}}_{j\sigma}
z ^{\vphantom{\dagger}}_{i\sigma})\ ,\nonumber
\end{eqnarray}
where $K=\langle c ^\dagger_i c ^{\vphantom{\dagger}}_j\rangle$
and $W=\langle z ^{\vphantom{\dagger}}_{i\sigma}
{\bar z} ^{\vphantom{\dagger}}_{j\sigma}\rangle$
Such a model was introduced by Sarker \cite{s96}, who identified a Curie
transition and found that the $e_g$\ fermion band becomes incoherent
above $T_ {\rm c}$.  For our purposes we are interested in phase separation.
Accounting for the possibility of condensation of Schwinger bosons, we write
$\Psi_{{\fam\mibfam\eightmib k}\sigma}\equiv\langle
z^{\vphantom{\dagger}}_{ {\fam\mibfam\eightmib k}\sigma}\rangle$.  Assuming
condensation only at $ {\fam\mibfam\tenmib k}=0$, we define
$\rho\equiv |\Psi_{ {\fam\mibfam\eightmib k}=0,\sigma}|^2$.
The mean field equations are then
\begin{eqnarray}
1+{1\over 2S}&=&\rho+{1\over 2S}\int_{-1}^1\!\!\!\!
d \gamma\,D( \gamma)\,
{\rm ctnh}\!\left({\Lambda-K\gamma\over 4S\Theta}\right)\nonumber\\
W&=&\rho+{1\over 2S}\int_{-1}^1\!\!\!\!
d \gamma\,\gamma\,D( \gamma)\,
{\rm ctnh}\!\left({\Lambda-K\gamma\over 4S\Theta}\right)\nonumber\\
\delta&=&{1\over 2}\int_{-1}^1\!\!\!\! d \gamma\,D( \gamma)\,
\tanh\!\left({\alpha +W \gamma \over 2\Theta}\right)\nonumber\\
K&=&{1\over 2}\int_{-1}^1\!\!\!\! d \gamma\, \gamma\,D( \gamma)\,
\tanh\!\left({\alpha +W \gamma\over 2\Theta}\right)\ ,
\label{sbmf}
\end{eqnarray}
where $\Lambda=\lambda/zt$, $\Theta= k_{ \scriptscriptstyle {\rm B}}T/zt$,
and $\alpha=\mu/zt$ ($B=zt$).

\begin{figure} [!t]
\centering
\leavevmode
\epsfxsize=8cm
\epsfysize=8cm
\epsfbox[18 144 592 718] {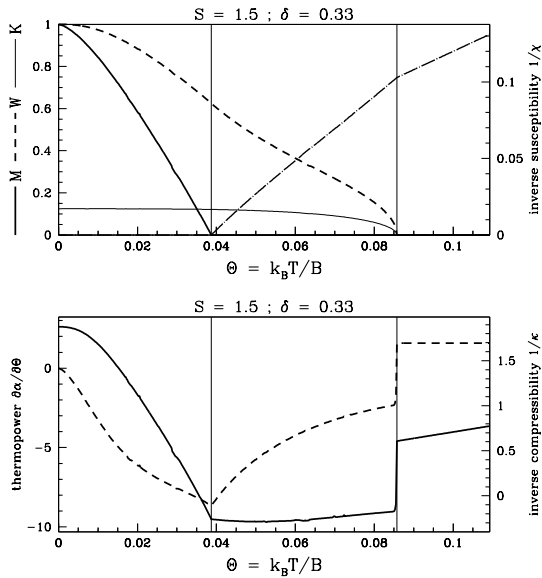}
\caption[]
{\label{fig2}
Solution to the model (\ref{hmf},\ref{sbmf}) using an elliptical
density of states and $S=\frac{3}{2}$ at $\delta=0.33$.
The bottom panel shows thermopower
(dashed; units of $k_{\scriptscriptstyle \rm B}$)
and inverse compressibility (solid), which is negative over
a band of temperatures.}
\end{figure}

We have solved the mean field equations for a semi-elliptic density of
states.  Several features of the solution are noteworthy.  Again as expected
there is a Curie transition at $\Theta_ {\rm c} (\delta)$.  In the absence of an
external magnetic field, the magnetization $M$ is equal to the condensate
fraction $\rho$ and vanishes for $\Theta>\Theta_ {\rm c} (\delta)$.
Second, the quantity $W$, which is directly proportional to the $c$
fermion bandwidth, is unremarkable through $T_ {\rm c}$.  It decreases
monotonically and eventually vanishes at a temperature
$\Theta_*(\delta)=(1+S^{-1})^{1/2}\sqrt{1-4\delta^2}/8\sqrt{2}$.
The dimensionless electronic kinetic energy $K$, which is proportional to
the boson bandwidth, also vanishes at this point.  The vanishing of $K$ and $W$
at $\Theta_*$ is likely a spurious artifact of the mean field theory;
it was noted in \cite{aa88,ag97}.  However, in a range $\Theta_ {\rm c}<\Theta
<\Theta_*$ above the Curie transition this theory describes a state with
vanishing magnetization yet finite and $\Theta$-dependent fermion bandwidth.
The mean field parameters are shown {\it versus\/} $\Theta$ in
Fig. \ref{fig2}.

Third, we find that $\kappa^{-1}(\Theta)$ has a discontinuity
in slope at $\Theta_ {\rm c}$ and a jump discontinuity at $\Theta_*$.
The locus of points where $\kappa^{-1}(\delta,\Theta)=0$ marks the
boundary of the region of phase separation.  This is shown in
Fig. \ref{fig1}(b).  Again corroborating our initial discussion, we find that
phase separation occurs for $|\delta|>\delta_0\simeq 0.289$, and for a range
of temperatures $\Theta\in [\Theta_ {\rm s} (\delta),\Theta_* (\delta)]$
surrounding $\Theta_ {\rm c}$.  The upper boundary
is $\Theta_*(\delta)$ itself; this is likely an artifact of the mean
field theory.  We expect that $W$ should continue to decrease as
$\Theta$ increases, tending to a finite asymptotic value that this
mean field approach cannot describe.
At any rate, for large $\Theta$ the third of the mean field equations
gives $\alpha\simeq \Theta\ln[(1+2\delta)/(1-2\delta)]$ and
$\kappa^{-1}\simeq 4\Theta/(1-4\delta^2)$.  Thus, $\kappa>0$ for
sufficiently large $\Theta$ and there is a finite range area of phase
separation.

The thermopower ${\cal S}$ is anomalous, owing to the dependence of
bandwidth on temperature through $W(\Theta)$.
It peaks at $\Theta_{\rm c}$ with a $\delta$-dependent value
${\cal S}_{\rm max}(\delta)$ which is enormous by standards of metal physics
(the dimensions of $\partial\mu/\partial T$ in Fig. \ref{fig2}
are restored by multiplying $\partial\alpha/\partial\Theta$ by
$k_{\scriptscriptstyle\rm B}/e=86.2\,\mu{\rm V}/{\rm K}$).
The order of magnitude of ${\cal S}$, and the existence of a peak 
near $T_{\rm c}$ agree with experiments\cite{za96}.
The sign of ${\cal S}$ is opposite to that predicted by a rigid band
model (for either carrier type).
At temperatures well above $\Theta_{\rm c}$, the temperature dependence
of the bandwidth vanishes, and the usual sign is restored.
Finally, at very low temperatures the thermopower also changes
sign -- the $T^{5/2}$ dependence of $1-W$ is eventually
overwhelmed by the $T^2$ dependence of $\mu$ coming from the Sommerfeld
expansion.  Within our model this happens at a very low
value of $T$ ({\it e.g.\/} $T/B=1.5\times 10^{-5}$ for $\delta=0.2$).

We have ignored so far all interactions except the double exchange mechanism.
Electrostatic effects will inhibit the full phase separation described here.
The standard RPA result ${\raise.35ex\hbox{$\chi$}}^{-1}(q)
={\raise.35ex\hbox{$\chi$}}_0^{-1}(q)-V(q)$, where
${\raise.35ex\hbox{$\chi$}}_0(q)$ is the bare charge susceptibility and
$V(q) = 4\pi e^2/q^2$, predicts an instability at finite wavevector
$q_*$ whenever $\lim_{q \rightarrow 0}{\raise.35ex\hbox{$\chi$}}_0(q)
=-{\cal V}^{-1}\partial x/\partial \mu$ is positive
(${\cal V}$ is the unit cell volume), with
\begin{equation}
q_*=\sqrt{-{4\pi e^2\over{\cal V}}\,{\partial x\over\partial\mu}}\ .
\end{equation} 
At low dopings, $q_*^2 \sim x$, while at higher fillings, $q_*$
is proportional to the Thomas-Fermi wavevector
$q_{\scriptscriptstyle{\rm TF}}$.
Thus, we expect that electrostatic effects will give rise 
to a domain structure, at a length scale of order $q_*^{-1}$.

In the CMR manganites, the $e_g$\ orbitals are doubly degenerate
in the absence of JT distortions.  While we expect intraatomic
Coulomb effects to preclude multiple occupancy of the $e_g$\ levels,
orbital ordering could significantly alter the picture discussed here
(for example, the sign of the thermopower could reverse).
We have also ignored interactions with the lattice. Our results suggest
that the tendency towards phase separation gives rise to domains with
a very low concentration of holes. In these domains, JT deformations
will be favored\cite{b98}. In general, the enhancement
of charge fluctuations near $T_ {\rm c}$ should induce significant
lattice deformations, because of the different sizes of the Mn$^{3+}$\ and
Mn$^{4+}$\ ions.  A structure with domains of different electronic
density and bandwidth enhances the tendency towards localization.
Standard models for the influence of critical spin fluctuations are
not applicable if, in addition, there are strong charge fluctuations.
A domain with extra charge and stronger ferromagnetic order is akin
to the spin polarons proposed in the literature.  In our case,
however, such a structure arises from magnetic effects alone. 

We are thankful to L. Brey, M. J. Calder\'on, and J. Fontcuberta
for useful discussions. F. G. was supported by CICyT (Spain) through
grant PB96-0875.


\begin{references}

\bibitem{wk55} E. D. Wollan and W. C. Koehler, {\sl Phys. Rev.}
{\bf 100}, 545 (1955).

\bibitem{g63} See, {\it e.g.\/}\ J. B. Goodenough, {\sl Magnetism
and the Chemical Bond} (Interscience, New York, 1963).

\bibitem{cvm97} J. M. D. Coey, M. Viret and S. von Molnar,
{\sl Adv. in Phys.} (in press, 1998).

\bibitem{z51} C. Zener, {\sl Phys. Rev.} {\bf 82}, 403 (1951).

\bibitem{mls95} A. J. Millis, P. B. Littlewood, and B. I. Shraiman,
{\sl Phys. Rev. Lett.} {\bf 74}, 5144 (1995).

\bibitem{rzb96}
J. Zang, A. R. Bishop and H. R\"oder,
{\sl Phys. Rev. B} {\bf 53}, R8840 (1996), 
Q. Li, J. Zang, A. R. Bishop and C. M. Soukoulis,
{\sl Phys. Rev. B} {\bf 56}, 4541 (1997).

\bibitem{v96}
C. M. Varma, {\sl Phys. Rev. B} {\bf 54}, 7328 (1996).

\bibitem{cb98}
M. J. Calderon and L. Brey, preprint ({\tt cond-mat/9801311}).

\bibitem{cvr95}
J. M. D. Coey, M. Viret, and L. Ranno,
{\sl Phys. Rev. Lett.} {\bf 75}, 3910 (1995).

\bibitem{ti97}
J. M. de Teresa {\it et al.\/},
{\sl Nature} {\bf 386}, 256 (1997).

\bibitem{msm96}
A. J. Millis, B. Shraiman and R. Mueller, {\sl Phys. Rev. Lett.}
{\bf 77}, 175 (1996).

\bibitem{g60} P.-G. de Gennes, {\sl Phys. Rev.} {\bf 118}, 141 (1960).

\bibitem{br70} W. F. Brinkman and T. M. Rice,
{\sl Phys. Rev. B} {\bf 2}, 1324 (1970).

\bibitem{n96} E. L. Nagaev, {\sl Physica B} {\bf 230-232}, 816 (1997).

\bibitem{dym97}J. Riera, K. Hallberg and E. Dagotto, {\sl Phys. Rev. Lett.}
{\bf 79}, 713 (1997),
S. Yunoki {\it et al.\/}, {\sl Phys. Rev. Lett.} {\bf 80}, 845 (1998).

\bibitem{ag97}
D. P. Arovas and F. Guinea, preprint ({\tt cond-mat/9711145}).

\bibitem{aa88} 
D. P. Arovas and A. Auerbach, {\sl Phys. Rev. B}
{\bf 38}, 316 (1988)

\bibitem{am76}
If we neglect the energy dependence of the scattering time, and
assume an isotropic Fermi velocity, the thermopower
is approximately given by ${\partial \mu/\partial T}$.  See,
{\it e.g.\/}\ N. W. Ashcroft and D. Mermin, {\it Solid State Physics},
Saunders Press, New York (1976).

\bibitem{ps96} W. E. Pickett and D. J. Singh, {\sl Phys. Rev. B} {\bf 53},
1146 (1996).

\bibitem{s96} 
S. K. Sarker, {\sl J. Phys.: Cond. Mat.} {\bf 8}, L515 (1996).

\bibitem{za96} J. Fontcuberta {\it et al.\/},
{\sl Appl. Phys. Lett.} {\bf 68}, 2288 (1996).

\bibitem{b98}
C. H. Booth, {\it et al.\/}
{\sl Phys. Rev. Lett.} {\bf 80}, 853 (1998).


\end{references}
\end{document}